\documentclass[12pt]{article}
\usepackage{graphicx}
\usepackage{latexsym,amsfonts}
\def\a{\alpha}
\def\b{\beta}

\def\d{\delta}

\def\m{\mu}
\def\n{\nu}

\def\s{\sigma}

\def\be{\begin{equation}}
\def\ee{\end{equation}}
\def\beq{\begin{eqnarray}}
\def\eeq{\end{eqnarray}}

\newcommand{\Cn}{{\mathbb{C}}}
\newcommand{\bqn}{\begin{eqnarray}}\newcommand{\eqn}{\end{eqnarray}}
\newtheorem{theorem}{Theorem}[subsection]

\begin{document}

\begin{titlepage}
\begin{flushright}
LPT-ENS/03-10 \\
ULB-TH-03/12 \\
\end{flushright}
\vskip 1.0cm

\begin{centering}

{\large {\bf Hyperbolic billiards of pure $D=4$ supergravities}}

\vspace{1cm}

Marc Henneaux$^{a,b}$
and
Bernard Julia$^{c}$ \\
\vspace{.7cm}
{\small
$^a$ Physique Th\'eorique et Math\'ematique,
Universit\'e Libre
de Bruxelles, C.P. 231, B-1050, Bruxelles, Belgium \\
\vspace{.2cm} $^b$ Centro de Estudios Cient\'{\i}ficos, Casilla
1469, Valdivia, Chile \\
\vspace{.2cm}
$^c$ Laboratoire de Physique Th\'eorique de l'Ecole Normale
Sup\'erieure, \\ 24, rue Lhomond, F-75231 Paris CEDEX 05, France}

\vspace{.5cm}

\end{centering}

\begin{abstract}
We compute the billiards that emerge in the
Belinskii-Khalatnikov-Lifshitz (BKL) limit for all pure
supergravities in $D=4$ spacetime dimensions, as well as for
$D=4$, $N=4$ supergravities coupled to $k$ ($N=4$) Maxwell
supermultiplets. We find that just as for the cases $N=0$ and
$N=8$ investigated previously, these billiards can be identified
with the fundamental Weyl chambers of hyperbolic Kac-Moody
algebras. Hence, the dynamics is chaotic in the BKL limit. A new
feature arises, however, which is that the relevant Kac-Moody
algebra can be the Lorentzian extension of a {\it twisted} affine
Kac-Moody algebra, while the $N=0$ and $N=8$ cases are untwisted.
This occurs for $N=5$, where one gets $A_4^{(2)\wedge}$, and for
$N=3$ and $2$, for which one gets $A_2^{(2)\wedge}$. An
understanding of this property is provided by showing that the
data relevant for determining the billiards are the {\it
restricted root system} and the {\it maximal split subalgebra} of
the finite-dimensional real symmetry algebra characterizing the
toroidal reduction to $D=3$ spacetime dimensions. To summarise:
split symmetry controls chaos.
\end{abstract}

\vfill
\end{titlepage}

\section{Introduction}
\setcounter{equation}{0} \setcounter{theorem}{0}
\setcounter{lemma}{0} As it has been shown recently, the classical
dynamics of the spatial scale factors and of the dilaton(s) (if
any) of $D$-dimensional gravity coupled to $p$-forms and scalar
fields can be described, in the vicinity of a spacelike
singularity, as a billiard motion in a region of hyperbolic space
bounded by hyperplanes \cite{DH3}\footnote{As it has become
standard practice in the field, the word {\em billiard} used as a
noun in the singular denotes the dynamical system consisting of a
ball moving freely on a ``table" (region in some Riemannian
space), with elastic bounces against the edges. {\em Billiard}
also sometimes means the table itself.}. This generalizes known
results for pure gravity in $D=4$ spacetime dimensions
\cite{BKL,Chitre,Misnerb}. Furthermore, in the case of the bosonic
sector of $11$-dimensional supergravity or $10$-dimensional
supergravities, the relevant billiard turns out to be identifiable
with the fundamental Weyl chamber of the Kac-Moody algebras
$E_{10}$, $BE_{10}$ or $DE_{10}$ \cite{DH3} -- which are
respectively the overextensions \cite{Julia,FF} of $E_8$, $B_8$
and $D_8$, i.e., $E_{10}\equiv E_8^{\wedge \wedge}$, $BE_{10}
\equiv B_8^{\wedge \wedge}$ and $DE_{10}\equiv D_8^{\wedge
\wedge}$ --, while for $D$-dimensional pure gravity, the algebra
is the overextension $A_{D-3}^{\wedge \wedge}$ of $A_{D-3}$
\cite{DHJN}. The geometrical reflexions occurring when the system
hits the billiard walls are fundamental Weyl reflexions and the
motion can thus be identified with an (infinite) Weyl word. The
fact that the underlying Kac-Moody algebras are hyperbolic
(provided $D<11$ for pure gravity) explains \cite{DHJN} the
chaotic behaviour of these systems as one approaches the
singularity \cite{DHS,DH1}.

When reduced to four spacetime dimensions, the above models
correspond to pure $N=8$ supergravity, or to $N=4$ supergravity
coupled to a collection of $N=4$ Maxwell/Yang-Mills multiplets
($6$ Maxwell multiplets for pure $N=4$, $D=10$ supergravity,
$6+16$ vector multiplets for $N=4$, $D=10$ supergravity with $E_8
\times E_8$ or $SO(32)$ Yang-Mills field). Only the $N=8$ case
defines a pure supergravity theory in four spacetime dimensions.
The purpose of this article is to investigate systematically the
billiards that describe the dynamics of all pure $D=4$
supergravities ($N= 1,2,3,4,5,6,8$) in the vicinity of a spacelike
singularity. We also investigate $D=4, N=4$ supergravity coupled
to a collection of an arbitrary number $k$ of $N=4$ vector
multiplets.

We find that the billiards for all these models can also be
associated with hyperbolic Kac-Moody algebras. Furthermore, we
prove that ``split symmetry controls chaos" in the following
sense: let ${\cal U}_3$ be the finite-dimensional real U-duality
algebra that appears in the toroidal compactification of the
theory to 3 dimensions. Then, the hyperbolic Kac-Moody algebra
whose fundamental Weyl chamber determines the billiard is the
overextension of the ``maximal split subalgebra" ${\cal F}$ of
${\cal U}_3$ (and not of ${\cal U}_3$ itself except when ${\cal
U}_3$ and ${\cal F}$ coincide, which occurs only when ${\cal U}_3$
is split, i.e., maximally non-compact). This explains in
particular why it is $BE_{10} \equiv B_8^{\wedge \wedge}$ (rather
than the overextension of the split form $D_{16}$ of
the non-split $so(8,8+16)$) that
determines the heterotic billiard.
However, while the Kac-Moody
algebras of the previously studied models are given by standard
overextensions of finite-dimensional Lie algebras, some of the
theories investigated here are characterized by a new feature: the
extension involves a ``twist". Specifically, the twist occurs for
$N=2,3$ and $5$, for which one gets respectively the Lorentzian
extensions $A_2^{(2)\wedge}$ and $A_4^{(2)\wedge}$ of the twisted
affine algebras $A_2^{(2)}$ and $A_4^{(2)}$ with Dynkin diagrams
\begin{figure}[ht]
\centerline{\includegraphics[scale=0.5]{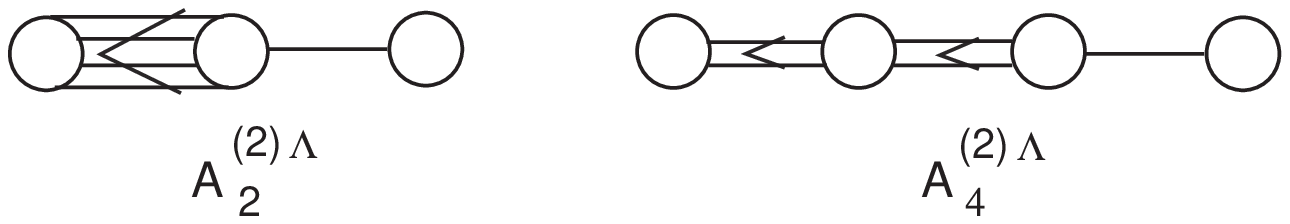}}
\end{figure}

\noindent
Note that the role of $B_8$ versus $so(8,8+16)$ was understood
in the context of non-split $U$-duality for the $so(8,8+r)$-models
in \cite{HJK}.  In that precise case, there is no twist.

In the next section, we set up our conventions and terminology.
Then, we derive the central theorem that relates the Kac-Moody
algebra ${\cal A}$ whose fundamental Weyl chamber is the billiard
table to the maximal split subalgebra ${\cal F}$ and the {\em
restricted} root system of the {\em real} symmetry algebra ${\cal
U}_3$ appearing in 3 spacetime dimensions (section
\ref{generaltheorem}). We show quite generally that ${\cal A}$ is
the overextension of ${\cal F}$, with a twist only when the root
system is of $bc$-type. The theorem covers all pure supergravity
models in $D=4$ spacetime dimensions, as well as $N=4, D=4$ SUGRA
with $k$ Maxwell multiplets. So when the U-duality algebra ${\cal
U}_3$ is not split (i.e. not maximally non-compact) its maximal
split subalgebra ${\cal F}$, which controls chaos, is smaller. The
analysis also explains the wall multiplicities. In section
\ref{N=2}, the billiard is computed in the particular case of
$N=2$ supergravity directly in $D=4$ dimensions, without going to
$D=3$ dimensions. This sheds a different light on the twist.
Finally, we relate the occurrence of the twist to a property of
real forms of untwisted but non split affine Kac-Moody algebras
\cite{VBV}.

\section{Conventions}
\setcounter{equation}{0} \setcounter{theorem}{0}
\setcounter{lemma}{0}

Let ${\cal G}_{\Cn}$ be a complex, finite-dimensional Lie algebra
and let ${\cal G}$ be one of its real forms. We denote by ${\cal
G}^{\wedge}$ the corresponding untwisted algebra of currents
(where all the currents are integer-moded). It is a real form of
the untwisted affine extension of ${\cal G}_{\Cn}$, defined from
the Chevalley-Serre presentation by adding generators associated
with the new, ``affine" root $\a_0= \d - \theta$, where $\theta$
is the highest root of ${\cal G}_{\Cn}$ and $\d$ the null root of
$({\cal G}_{\Cn})^{\wedge}$ (see \cite{Kac,MP}; $({\cal
G}_{\Cn})^{\wedge}$ is ${\cal G}_{\Cn}^{(1)}$ in Kac's notations).
We shall also consider real twisted affine algebras, but only in
their split form. These are again defined from the standard Dynkin
diagrams in terms of Chevalley-Serre generators and relations, but
one considers only real combinations. We adopt Kac's notations
\cite{Kac} in the twisted case. Given a split (untwisted or
twisted) affine Kac-Moody algebra ${\cal E}$, we define its real
Lorentzian extension ${\cal E}^{\wedge}$ by adding an
``overextended root" \cite{Julia,FF} to its Dynkin diagram and
considering the corresponding set of generators and relations over
the reals. The overextended root is attached to the affine root
$\alpha_0$ with a single line in the untwisted case. For the
twisted algebras $A_{2\ell}^{(2)}$ -- the only twisted cases we
shall encounter here --, the overextended root is attached to the
unique root carrying label $1$, which is the longest root. Note
that this longest root is {\em not} the root $\alpha_0$ of
\cite{Kac}.  More information on overextensions is given in the
appendix.

To characterize the real form ${\cal G}$, we adopt the Tits-Sakate
theory as developed for instance in \cite{Araki}. One selects a
maximally non-compact Cartan subalgebra of ${\cal G}$ and one
diagonalizes simultaneously in ${\cal G}$ all the operators $ad_h$
with $h$ in this Cartan subalgebra. This defines the restricted
roots and the restricted root system $\bar B$, which can be either
one of the standard reduced root systems $a_n$, $b_n$, $c_n$,
$d_n$, $g_2$, $f_4$, $e_6$, $e_7$ or $e_8$, or one of the nonreduced
$bc_n$ systems. The $bc_n$ root system is obtained by merging the
$b_n$ and $c_n$ root systems in such a way that the long roots of
$b_n$ are the short roots of $c_n$. It is called nonreduced
because $\a$ and $2 \a$ can be simultaneously roots (if $\a$ is a
short root of $b_n$, $2 \a$ is a long root of $c_n$). The
restricted roots include all the standard roots of a Lie
algebra, which we call the ``maximal split subalgebra" following Borel
and Tits \cite{BorelTits} and denote
by ${\cal F}$. Note, however, that there are in general
crucial differences between the restricted root
system of ${\cal G}$ and the standard root system of ${\cal F}$ :
(i) twice a root of ${\cal F}$ can be a restricted root
and (ii) the restricted roots can come with a non trivial
multiplicity. The dimension of the maximally non-compact Cartan
subalgebra of ${\cal G}$, which is the Cartan subalgebra of ${\cal
F}$, is called the real rank of ${\cal G}$ and denoted by $r$. It
is only if ${\cal G}$ is the split real form of ${\cal G}_\Cn$
that ${\cal F}$ coincides with ${\cal G}$. This information is
encoded in the Tits-Sakate diagrams and given for
instance in \cite{Helgason} (table VI, chapter X). The importance of
non split real forms of U-duality groups dates back to supergravity days
\cite{Julia}. The facts that its precise real form is related to the
$SL(D-d,R)$ compactification symmetry and that this fixes the maximal
oxidation possible have also been used and exposed repeatedly.

\section{A general theorem}
\setcounter{equation}{0} \setcounter{theorem}{0}
\setcounter{lemma}{0} \label{generaltheorem}

When reduced to $D=3$ spacetime dimensions, the bosonic sectors of
all $D=4$ pure supergravity theories become the Einstein theory
coupled to a coset model $G / H$ where $H$ is the maximal compact
subgroup of $G$, and where $G$ depends on $N$ \cite{BJ1}. We give
in the first column of table I of the concluding section the
(real) Lie algebra ${\cal G} \equiv {\cal U}_3$
of the $D=3$ symmetry group $G$. The
information is taken from \cite{BJ1}.

Now, the computation of the billiard can be carried out in any
number of dimensions $\geq 3$ because the dominant walls that
define the billiard are invariant under toroidal dimensional
reduction and dualization \cite{DdBHS}. If the theory is
explicitly known in $3$ dimensions, which is the case here, the
easiest way to uncover the underlying Kac-Moody structure is to
analyze the billiard in that dimension, where it is particularly
transparent. In this section, we determine the billiard for
general coset models coupled to gravity in $3$ dimensions, where
${\cal G}$ is not necessarily the split real form of ${\cal
G}_\Cn$. Our theorem will be a generalization to the non-split
case of the results obtained in \cite{DdBHS} for the split case,
where the Lagrangians of \cite{BGM,CJLP} of maximally non-compact
coset models were investigated.

The basic tool will be the Iwasawa decomposition of the non
compact group $G$ where the restricted root system plays a central
r\^ole (see e.g. \cite{Helgason}, chapter IX). Using this
decomposition, the $G/H$ coset action takes the form \beq
&&S^{\phi^\a, \chi^A} =
-\int g^{\m \n} \sqrt{-g}\big( \sum_\a
\partial_\m \phi^\a
\partial_\n \phi^\a \nonumber \\
&& \hspace{2cm} + \frac{1}{2} \sum_A e^{2 \lambda^A(\phi)}
(\partial_\m \chi^A + \cdots ) (\partial_\n \chi^A +
\cdots) \big) d^3x
\label{iwa}
\end{eqnarray}
where the ellipsis denote ``correction terms" to the ``abelian
curvatures" $d\chi^A$. In (\ref{iwa}), the $\phi^\a$'s are the
dilatons, whose number is equal to the real rank $r$ of ${\cal
G}$. The dilatons can be identified with coordinates on the
maximally non-compact Cartan subalgebra of ${\cal G}$. The linear
forms $\lambda^A(\phi)$ are the restricted roots (taking into
account multiplicities, i.e., if a restricted root has
multiplicity $k$, there are then $k$ linear forms
$\lambda^A(\phi)$ associated with it, say $\lambda^1(\phi), \cdots
, \lambda^k(\phi)$, which are equal, $\lambda^1(\phi) =
\lambda^2(\phi) = \cdots = \lambda^k(\phi)$). There is an axion
field $\chi^A$ for each linear form $\lambda^A(\phi)$. We shall
denote by $\theta(\phi)$ the highest root of the restricted root
system, hereafter called $\bar{B}$.

The complete action for the system, including gravity, is the sum
of (\ref{iwa}) and of the Einstein action \be S^E = \int
\sqrt{-g}\, R \, d^3x, \ee i.e., \be S = S^E + S^{\phi^\a, \chi^A}
\label{3daction}\ee As in \cite{DH1}, we normalize the dilaton
kinetic term such that it has weight one with respect to the
Einstein term. To get the billiard walls, one decomposes the
$2$-dimensional spatial metric $g_{ij}$ as\footnote{The rules for
writing down the billiards have been stated in
\cite{DH1,DH3,DHJN}. A systematic derivation is presented in
\cite{DHN}.}
\begin{equation}
g_{ij} = \left( \matrix{e^{-2 \b^1} &n e^{-2 \b^1} \cr n e^{-2
\b^1} & n^2 e^{-2 \b^1} + e^{-2 \b^2} \cr } \right)
\label{matrix}
\end{equation}
We shall call collectively ``(logarithmic) scale factors" both the
$\b^i$'s ($i=1,2$) and the dilatons $\phi^\a$. The action
determines a metric in the space of the scale factors which reads,
in our normalization \be d\s^2 = \sum_{i=1}^2 (d \b^i)^2 -
(\sum_{i=1}^2 d\b^i)^2 + \sum_{\a= 1}^r (d\phi^\a)^2 \ee The
inverse metric is \be (\partial f \vert \partial f) = \sum_{i=1}^2
(\partial_i f)^2 - (\sum_{i=1}^2
\partial_i f)^2 + (\sum_{\a=1}^r
\partial_\a f)^2
\ee The normalization of the roots of the restricted root system
is such that the highest root $\theta$ has length squared equal to
$2$, \be (\theta \vert \theta) = 2 .\ee The reason for this will
be given below.

The linear wall forms defining the billiard associated with the
action (\ref{3daction}) are the following \cite{DH3,DHN}
\begin{itemize}
\item Axion electric walls, coming from the kinetic energy
$(\dot{\chi}^A)^2$ of the axions \be w^A_E = \lambda^A(\phi)
\label{electric} \ee \item Axion magnetic walls, coming from the
potential energy $(\partial_k \chi^A)^2$ of the axions \be
w^{A,i}_M = \b^i - \lambda^A(\phi) \label{magnetic}\ee \item
Symmetry wall, coming from the kinetic term of the Iwasawa
parameter $n$ in (\ref{matrix}) \be w_S = \b^2 - \b^1
\label{symmetry}\ee
\end{itemize}
The same electric or magnetic wall forms may occur several times,
but this does not affect the analysis. The billiard is defined by
the inequalities $w_\Gamma \geq 0$ where $\Gamma$ runs over all
wall forms (if one inequality occurs several times, we clearly
only need to keep it once, which is why multiplicities of wall
forms are not important here). The walls are $w_\Gamma = 0$. Since
some of the wall forms can be expressed as linear combinations
with non-negative (integer) coefficients of a smaller subset of
wall forms, only this smaller subset is relevant. The relevant
subset is easily determined to contain
\begin{enumerate}
\item The electric wall form \be \s_\a (\phi)
\label{dominantelectric}\ee where the $\s_\a$ are the simple roots
of the restricted root system $\bar B$ \item The magnetic wall
form \be \b^1 - \theta(\phi)\label{dominantmagnetic}\ee where
$\theta(\phi)$ is the highest root of $\bar B$ \item The symmetry
wall (\ref{symmetry}).
\end{enumerate}
We denote collectively the dominant wall forms by $\a_i$ ($i=1,
\cdots, r, r+1, r+2$).

The dominant wall forms are identified with the simple roots of
the searched-for Kac-Moody algebra \cite{DH3,DHJN}. Once the
dominant wall forms have been determined, one simply computes the
Cartan matrix of the Kac-Moody algebra through the familiar
formula \be A_{ij} = 2 \frac{(\a_i \vert \a_j)}{(\a_i \vert \a_i)}
\label{Cartan}\ee The following theorem is a straightforward
consequence of the previous considerations and constitutes the
core result of our paper.
\begin{theorem} (i) If the restricted root system is not of
$bc$-type, the billiard is the fundamental Weyl chamber of the
Kac-Moody algebra ${\cal F}^{\wedge \wedge}$ (overextension of the
maximal split subalgebra ${\cal F}$);

(ii) If the restricted root system is $bc_r$, the billiard is the
fundamental Weyl chamber of the Kac-Moody algebra
$A_{2r}^{(2)\wedge}$.
\end{theorem}

\vspace{.4cm} {\bf Proof:} One must compute the Cartan matrix.
\begin{itemize}
\item In the first step, we determine the ``electric" submatrix of
the Cartan matrix obtained by restricting (\ref{Cartan}) to the
dominant electric wall forms (\ref{dominantelectric}). This
clearly yields the Cartan matrix of the maximal split subalgebra
${\cal F}$ since the $\s_\a$ are the simple roots. \item We next
add the dominant magnetic wall form (\ref{dominantmagnetic}).
\begin{itemize} \item If the restricted root system is not of $bc$-type, the
highest root $\theta(\phi)$ is also the highest root of the root
system of ${\cal F}$. Since the linear form $\b^1$ has length
squared equal to zero and is orthogonal to any linear form
involving only the dilatons, we see that the dominant magnetic
wall form (\ref{dominantmagnetic}) has length squared equal to two
and is such that $(\b^1 - \theta \vert \s_\a) = - (\theta \vert
\s_\a)$. This enables one to identify the dominant magnetic wall
form with the affine root of the untwisted affine extension ${\cal
F}^{\wedge}$ of ${\cal F}$ \cite{Kac,MP}. At this stage, we thus
have ${\cal F}^{\wedge}$. \item If the restricted root system is
of $bc$-type, say $bc_r$, the first step yields the Cartan matrix
of $b_r$. We order the simple roots of $\bar B$ so that $\s_1$ is
the short root and $\s_i$ is linked to $\s_{i-1}$ and $\s_{i+1}$
($1<i<r$). The highest root $\theta$ of the $bc_r$-system is
connected only to $\s_r$ and its length squared is four times that
of a short root (\cite{Helgason}, chapter X). This yields the
Cartan matrix of the twisted affine algebra $A_{2r}^{(2)}$ (which
one might in fact denote, with adapted conventions,
$BC_r^{\wedge}$, as in \cite{MP}). Note that since the highest
root has squared length equal to $2$, the short root $\s_1$ has
squared length $1/2$, the other simple roots having squared length
equal to $1$.
\end{itemize}
\item What remains to be done is to add the symmetry wall. The
only non-vanishing scalar products involving $w_S$ are $(w_S \vert
w_S) = 2$ and $(w_S \vert \b^1 - \theta) = -1$. Thus $w_S$ is
attached by a single line to the dominant magnetic root $\b^1 -
\theta$. This yields in all cases the Lorentzian extension of the
affine algebra obtained in the previous step, the overextended
root being $w_S$. Thus we do indeed get ${\cal F}^{\wedge \wedge}$
if the restricted root system is not of $bc$-type, and
$A_{2k}^{(2)\wedge}$ otherwise. $\Box$
\end{itemize}

In the particular case when ${\cal G}$ is the split real form of
${\cal G}_\Cn$, one has ${\cal G} = {\cal F}$ and the restricted
root system coincides with the root system of ${\cal G}$ (which is
reduced and for which each root has multiplicity one). The
relevant Kac-Moody algebra is then ${\cal G}^{\wedge \wedge}$ as a
particular case of (i) in the theorem. This result has been
established previously in \cite{DdBHS} along similar
$3$-dimensional lines. The reason that $(\theta \vert \theta) = 2$
is also the same as in that paper. It comes from the fact that the
theory is the reduction of a higher-dimensional one. Indeed, the
dominant magnetic wall (\ref{dominantmagnetic}) is a symmetry wall
in higher dimensions\footnote{In spacetime dimension $D$, the
dominant walls include always the symmetry walls $\b'^i - \b'^j$
($i>j$) where $\b'^i$ are the scale factors in $D$ dimensions. If
$D= d+1>3$ is the endpoint of the oxidation sequence, $\b'^d -
\b'^{d-1}$ becomes upon dimensional reduction the symmetry wall
(\ref{symmetry}) (the indices $(1,2)$ in $3$ dimensions correspond
to $(d-1,d)$ in $D$ dimensions) \cite{DdBHS} while
$\b'^{d-1}-\b'^{d-2}$ becomes a wall of the form $\b^1 - k(\phi)$
with $k(\phi)$ some linear form in the dilatons. Since this wall
is dominant and since the only dominant wall in $3$ dimensions of
the form $\b^1 - k(\phi)$ is the dominant magnetic wall
(\ref{dominantmagnetic}), they must indeed be equal.} and these
walls have all length squared equal to two \cite{DH3}. This
normalization of the roots of the coset space might not hold for
theories that cannot be oxidized \cite{CJLP} (see also \cite{Keur}
for a group-theory approach to oxidation). If the normalization of
$\theta$ were changed, one would not get the Kac-Moody algebras
listed below.

Since all the supergravity models under consideration fall within
the scope of the above theorem, it is now immediate to determine
the associated billiard Kac-Moody algebra. The results are
collected in table I of the concluding section. Note, as a further
check, that the highest root $\theta(\phi)$ is always non
degenerate for these models (see table VI in chapter X of
\cite{Helgason}). This is necessary because we have seen that
$\b^1 - \theta$ is a symmetry wall in higher dimensions, and
symmetry walls are never degenerate \cite{DHN}. A highest root
with multiplicity $>1$ is thus an obstruction to oxidation. Note
also that all the algebras ${\cal A}$ of the table are hyperbolic,
so that the models are all chaotic.

\section{$N=2$ pure supergravity}
\setcounter{equation}{0} \setcounter{theorem}{0}
\setcounter{lemma}{0} \label{N=2}

In order to understand better the twist, we shall repeat the
billiard computation in $D=4$ spacetime dimensions for the
simplest case where the twist is present, namely, $N=2$ pure
supergravity. The bosonic sector is then the Einstein-Maxwell
theory with one Maxwell field. Following the rules of
\cite{DH3,DH1}, one gets the following billiard wall forms
$w_\Gamma$ ($\Gamma=1, \cdots, 12$),
\begin{itemize}
\item Symmetry wall forms \be \b'^2- \b'^1 , \; \; \; \b'^3 -
\b'^2 , \; \; \; \b'^3 - \b'^1\ee \item Curvature wall forms \be
2\b'^1 , \; \; \; 2 \b'^2 , \; \; \; 2 \b'^3 \ee \item Electric
wall forms \be \b'^1 , \; \; \; \b'^2 , \; \; \;
\b'^3 \ee
\item Magnetic wall forms \be \b'^1 , \; \; \; \b'^2 , \; \; \;
\b'^3 \ee
\end{itemize}
where the $\b'^i$ are the (logarithmic) scale factors in $D=4$
spacetime dimensions.

The billiard is clearly defined by the subset of inequalities \be
\a_1(\b') \equiv \b'^1 \geq 0 , \; \; \; \a_2(\b') \equiv \b'^2-
\b'^1 \geq 0 , \; \; \; \a_3(\b') \equiv \b'^3 - \b'^2 \geq
0.\label{leading}\ee since the other inequalities are obvious
consequences of this subset. A straightforward calculation, using
the metric in the space of the scale factors, \be \sum_{i=1}^3 (d
\b'^i)^2 - (\sum_{i=1}^3 d\b'^i)^2 \ee or more properly, its
inverse \be \sum_{i=1}^3 (\partial_i f)^2 - \frac{1}{2}
(\sum_{i=1}^3
\partial_i f)^2 \ee shows that the matrix
$$ A_{ij} = 2 \frac{(\a_i \vert \a_j)}{(\a_i \vert \a_i)} $$ is
equal to \be A_{ij} = \left( \begin{array}{ccc}
2 & -4 & 0 \\
-1 & 2 & -1 \\
0 & -1 & 2 \\
\end{array} \right) \label{A22}\ee
The $2 \times 2$ submatrix \be \left( \begin{array}{ccc}
2 & -4 \\
-1 & 2 \\
\end{array} \right) \ee
is the Cartan matrix of $A_2^{(2)}$ \cite{Kac}, the second root
$\a_2$ being the long root. The third root $\a_3$ is clearly
attached to $\a_2$ with a single line, which enables one to
identify (\ref{A22}) with the Cartan matrix of the Lorentzian
extension $A_2^{(2)\wedge}$ of the twisted affine algebra
$A_2^{(2)}$.

If instead of the Einstein-Maxwell theory, we had the pure
Einstein theory, the electromagnetic electric and magnetic walls
would be absent and the billiard would be defined by the subset of
inequalities \be \tilde{\a}_1(\b') \equiv 2 \b'^1 \geq 0 , \; \;
\; \a_2(\b') \equiv \b'^2- \b'^1 \geq 0 , \; \; \; \a_3(\b')
\equiv \b'^3 - \b'^2 \geq 0\label{leading'}\ee instead of
(\ref{leading}), $\tilde{\a}_1$ being the dominant curvature wall
form. Both (\ref{leading'}) and (\ref{leading}) define the same
billiard but the normalization of the first root, which is an
information contained in the Lagrangian \cite{DH3}, is different.
The Cartan matrix associated with (\ref{leading'}) is \be
\tilde{A}_{ij} = \left( \begin{array}{ccc}
2 & -2 & 0 \\
-2 & 2 & -1 \\
0 & -1 & 2 \\
\end{array} \right) \label{A1over}\ee This is the Cartan matrix
of $A_1^{\wedge \wedge}$ \cite{DHJN}. The passage from
$A_1^{\wedge \wedge}$ to $A_2^{(2)\wedge}$ when one includes the
Maxwell field comes from the fact that the curvature root
$\tilde{\a}_1$, which is dominant in the absence of
electromagnetism, ceases to be so and is replaced by $\a_1$
($\tilde{\a}_1 = 2 \a_1$). The two algebras have clearly the same
fundamental Weyl chamber and Weyl group.

We also see directly on the $4$-dimensional formulation that the
short root $\a_1$ is twice degenerate, it appears once as electric
wall form and once as magnetic wall form (because of
electric-magnetic duality, the electric and magnetic energy
densities contribute the same wall forms). Furthermore, twice the
root $\a_1$ is also a root (in fact, a curvature root), another
feature characteristic of the non reduced restricted root system
of $bc_1$-type. This matches of course perfectly the computation
in $3$ spacetime dimensions. When reduced to $D=3$ spacetime
dimensions, the $D=4$ Einstein-Maxwell system yields a $SU(2,1)$
coset model ($SU(2,1)/S(U(2) \times U(1))$) coupled to gravity
\cite{BJ1}. The real rank is one and there is only one dilaton
$\phi$. The restricted root system of the coset is of $bc_1$-type
\cite{Helgason}. By applying the standard formulas of dimensional
reduction, one finds that the reduced action has indeed the form
of (\ref{3daction}), where the restricted roots are
$\frac{1}{\sqrt{2}} \phi$ (degenerated twice) and $\sqrt{2} \phi$
(highest root with multiplicity one). The $3D$ electric wall is
just the $4D$ electric (or magnetic) wall $\a_1$ in
(\ref{leading}); the $3D$ magnetic wall is the $4D$ symmetry wall
$\a_2$ in (\ref{leading}); and the $3D$ symmetry wall is the $4D$
symmetry wall $\a_3$ in (\ref{leading}): everything matches once
the $3D \rightarrow 4D$ translation is appropriately done.

\section{Twisted case - link with real forms of affine algebras}
\setcounter{equation}{0} \setcounter{theorem}{0}
\setcounter{lemma}{0}

As we have seen, adding the last overextended root is direct and
involves no subtlety. The twist arises already - and only - at
the level of the affinization. With this in mind, the appearance
of twisted algebras in the description of the billiard is not
surprizing if one recalls the theory of real forms of affine
Kac-Moody algebras. For the ``almost split'' case their
classification is given in \cite{VBV}. By mere inspection of the
tables given in that paper, it can be observed that if ${\cal F}$
is the maximal split subalgebra of the finite-dimensional real Lie
algebra ${\cal G}$, the affine extension ${\cal F}^\wedge$ {\em
may not be the maximal split subalgebra of} ${\cal G}^\wedge$. In
particular, the maximal split subalgebra of the current algebras
$su(2,1)^\wedge$ and $su(4,1)^\wedge$ is $A_2^{(2)}$, while the
maximal split subalgebra of the current algebra $E_{6 \vert
-14}^\wedge$ is $A_4^{(2)}$. Note that for both $A_2^{(2)}$ and
$A_4^{(2)}$, twice the shortest simple root is a root, a feature
that the authors of \cite{VBV} denote by adding a $\times$ over
the short root on the restricted Dynkin diagram.

On the other hand, the maximal split subalgebras of the other
current algebras relevant to the supergravity models under
consideration, namely, $A_1^\wedge$, $so(8,k+2)^\wedge$ (with $0
\leq k <6$), $so(8,8)^\wedge$, $so(8,k+2)^\wedge$ (with $k>6$),
$E_{7 \vert -5}^{\wedge}$ and $E_{8 \vert +8}^\wedge$ involve no
twist and are respectively $A_1^\wedge$, $B_{k+2}^\wedge$,
$D_8^\wedge$, $B_8^\wedge$, $F_4^\wedge$ and $E_8^\wedge \equiv
E_{8 \vert +8}^\wedge$.

\section{Conclusions and summary}
\setcounter{equation}{0} \setcounter{theorem}{0}
\setcounter{lemma}{0}

In this paper, we have established a theorem that enables one to
derive the billiard dynamics of all $D=4$ pure supergravity models
directly from their $D=3$ formulation: the relevant data that
control the dynamics in the vicinity of a spacelike singularity
are the restricted root system and the maximal split subalgebra of
the real $D=3$ symmetry algebra. The precise rule is: the
Kac-Moody algebra ${\cal A}$ relevant for the billiard motion is
just the overextension of the maximal split subalgebra ${\cal F}$,
${\cal A} = {\cal F}^{\wedge \wedge}$, except when the restricted
root system is of $bc$-type, in which case there is a twist.
Because the Kac-Moody algebras that emerge for the models are all
hyperbolic, the dynamics is asymptotically chaotic.

Our theorem goes beyond $D=4$ pure supergravities and covers all
systems whose reduction to $D=3$ is a non-linear sigma model $G/H$
coupled to gravity. This is the case for $D=4$ supergravities
with $k$ vector (``Maxwell") multiplets, for which the $D=3$
symmetry algebra is $so(8,k+2)$. It is also the case, for
instance, for the $N=2$ $D=5$ exceptional Einstein-Maxwell
theories of \cite{Guna1,Guna2}. Since these latter models are
described in $D=3$ by the real Lie algebras $F_{4 \vert 4}$, $E_{6
\vert 2}$, $E_{7 \vert -5}$ or $E_{8 \vert -24}$ (corresponding to
the real, complex, quaternionic or octonionic Jordan algebras,
respectively) \cite{Guna2}, one can immediately infer that their
billiard is the fundamental Weyl chamber of the hyperbolic
Kac-Moody algebra $F_4^{\wedge \wedge}$. Indeed, $F_4$ is in all
cases the maximal split subalgebra (and the restricted root system
is of $f_4$ type).

Our results are summarized in the following table, in which we
give the number of supersymmetries ($N$), the real symmetry
algebra that emerges in $D=3$ dimensions (${\cal U}_3$), the
corresponding restricted root system ($\bar{B}$), the maximal
split subalgebra (${\cal F}$) and the hyperbolic Kac-Moody algebra
that controls the BKL limit (${\cal A}$). In the last three lines,
$k$ is the number of coupled Maxwell multiplets (thus the fourth
line corresponds to $k=0$). We denote the root systems by small
letters to distinguish them from the Lie algebras.

$$\begin{array}{ccccc}\vspace{.3cm} \hspace{2.5cm} & & \hbox{{\bf D=4 SUGRAS}}&
& \hspace{2.5cm}\\
\vspace{.3cm} N & {\cal U}_3 & {\bar B} & {\cal F}& {\cal A}
\\ N=1
& sl(2,R)& a_1 & A_1& A_1^{\wedge \wedge}\\
N=2 & su(2,1)& bc_1& A_1 & A_2^{(2)\wedge} \\
N=3 & su(4,1) & bc_1&A_1 & A_2^{(2)\wedge} \\ N=4 & so(8,2) &
c_2& C_2 &C_2^{\wedge \wedge} \\ N=5 & E_{6\vert-14} & bc_2 &C_2&
A_4^{(2)\wedge}\\N=6 & E_{7 \vert -5} & f_4& F_4 & F_4^{\wedge
\wedge}
\\
N=8 & E_{8 \vert + 8} & e_8 &E_8 & E_8^{\wedge \wedge}\\
N=4, \; k < 6 & so(8,k+2) &b_{k+2}&B_{k+2} &B_{k+2}^{\wedge \wedge} \\
N=4, \; k=6 & so(8,8) & d_8&D_8 & D_8^{\wedge \wedge} \\
N=4, \; k>6 & so(8,k+2)& b_8&B_8 & B_8^{\wedge \wedge} \vspace{.3cm}\\
\vspace{.3cm} \hspace{2.5cm} & & \hbox{{\bf TABLE I}}&
& \hspace{2.5cm}\\
\end{array}$$

\noindent [Recall the equivalences $A_1 \equiv B_1 \equiv C_1$ and
$B_2 \equiv C_2$.] The real algebra ${\cal F}$ is by definition
split, i.e., it always corresponds to the maximally non-compact
real form, which is for that reason not explicitly written (so, in
the ${\cal F}$-column, $A_1 \equiv A_{1 \vert + 1} \equiv
sl(2,R)$, $B_8 \equiv so(8,9)$ etc). Similarly, the Kac-Moody
algebra ${\cal A}$ in the last column is split (real linear
combinations of the Chevalley generators and of their multiple
commutators).

The restricted root system is of $bc$-type only for $N=2, 3$ and
$N=5$. There is then a twist. For $N=2$ and $N=3$, one gets
$A_2^{(2) \wedge}$ instead of $A_1^{\wedge \wedge}$; while for
$N=5$, it is $A_4^{(2) \wedge}$ that appears rather than
$C_2^{\wedge \wedge}$. Note that the actual twisted algebra that
emerges has not only the same rank as the standard (untwisted)
overextension ${\cal F}^{\wedge \wedge}$,
but also the same Weyl group, \beq W(A_1^{\wedge \wedge}) &\simeq&
W(A_2^{(2) \wedge}),
\\ W(C_2^{\wedge \wedge}) &\simeq& W(A_4^{(2) \wedge})\eeq

One of the interests of the billiard analysis is its connection
with $U$-dualities \cite{BP} and hidden symmetries of the theory,
for which various proposals exist
\cite{BJ1,PW,BJall1,DHN2,BJall2,En}. We reserve for further
study a more detailed analysis of the significance of the twist in
the symmetry structure of the models where it appears.

Finally, we note that although derived with the purpose of
determining the billiard structure of supergravity theories,
our theorem makes no use of supersymmetry.  As stressed
previously, the only relevant datum
is the real symmetry group ${\cal U}_3$
which characterizes the
manifold of the scalar fields coupled to gravity in the toroidal
compactification of the theory to three dimensions.

\section*{Acknowledgements}
M.H. is grateful to Thibault Damour and Hermann Nicolai for many
fruitful discussions and to Christiane Schomblond for
technical help, and B.J. to V. Kac and G. Rousseau for
useful references. MH also thanks the ``Laboratoire de Physique
Math\'ematique et Th\'eorique" of the University of
Montpellier for kind hospitality while this article was finished.
The work of M.H. is supported in part by the ``Actions de
Recherche Concert\'ees", a "P\^ole d'Attraction
Interuniversitaire" (Belgium) and by IISN-Belgium (convention
4.4505.86). Support from the European Commission RTN programme
HPRN-CT-00131 is also gratefully acknowledged.

\section*{Appendix A:
Overextensions of \\ finite-dimensional simple Lie algebras}

Let ${\cal L}$ be a complex, finite-dimensional, simple Lie
algebra of rank $r$, with simple roots $\alpha_1, \alpha_2,
\cdots, \alpha_r$.  We normalize the roots so that the long roots
have squared length equal to $2$ (the short roots, if any, have then
squared length equal to $1$ (or $2/3$ for $G_2$)). The roots
of simply-laced algebras are regarded as long roots.

We denote by
$\theta$ the highest root.  It is a long root.
We denote by $V$ the $r$-dimensional Euclidean vector
space spanned by $\alpha_i$ ($i= 1, \cdots, r$).
Let $M_2$ be the 2-dimensional Minkowski space with
basis vectors $u$ and $v$ so that $(u \vert u) =
(v \vert v) = 0$ and $(u \vert v) = 1$.  The
metric in the space
$V \oplus M_2$ has clearly Minkowskian signature $(-, +, +,
\cdots, +)$ so that any Kac-Moody algebra whose simple roots
span $V \oplus M_2$ is necessarily Lorentzian.

\subsection*{Standard overextensions}
The standard overextensions ${\cal L}^{\wedge \wedge}$
are obtained by adding to the original roots of
${\cal L}$ the roots
$$ \alpha_0 = u - \theta, \; \; \; \; \alpha_{-1}
= -u - v $$
The root $\alpha_0$ is called the affine root and the
algebra ${\cal L}^\wedge (\equiv {\cal L}^{(1)})$
with roots $\alpha_0, \alpha_1,
\cdots, \alpha_r$ is the untwisted affine extension of
${\cal L}$.  The root $\alpha_{-1}$ is known as the
overextended root.  One has clearly rank$({\cal L}^{\wedge \wedge})
=$ rank$({\cal L}) +2$.

The algebras $A_k^{\wedge \wedge}$ ($k \leq 7$),
$B_k^{\wedge \wedge}$ ($k \leq 8$), $C_k^{\wedge \wedge}$ ($k \leq
4$), $D_k^{\wedge \wedge}$ ($k \leq 8$), $G_2^{\wedge \wedge}$
$F_4^{\wedge \wedge}$, $E_k^{\wedge \wedge}$ ($k=6,7,8$) are
hyperbolic.  [The list of hyperbolic KM algebras may be found in
\cite{Sa}.]

\subsection*{Twisted overextensions}
Twisted affine algebras are related to either
the $bc$-root systems or to extensions by the highest
short root (see \cite{Kac}, proposition 6.4).

\subsubsection*{Twisted overextensions associated with
the $bc$-root systems}
These are the overextensions met in the text.  The construction
proceeds as for the untwisted overextensions, but
the starting point is now the $bc_r$ root system. The restricted
Dynkin diagram of $bc_r$ is the Dynkin diagram of $B_r$ with a
$\times$ over the simple short root, say $\alpha_1$,
to indicate that $2 \alpha_1$ is also a root.
The roots are also rescaled by the factor $(1/\sqrt{2})$
so that the highest root $\theta$ of the
$bc$-system has length 2 (instead of 4).  Indeed,
$\theta$ is given by $\theta =
2(\alpha_1 + \alpha_2 + \cdots + \alpha_r)$
\cite{Helgason}. It has squared length
equal to 2 (with the rescaling)
and has non-vanishing scalar product only with
$\alpha_r$ ($(\alpha_r \vert \theta) = 1$). The overextension
procedure yields the algebra $BC_r^{\wedge \wedge} \equiv
A_{2 r}^{(2)\wedge}$.

There is an alternative overextension $A_{2 r}^{(2)'\wedge}$
that can be defined
by starting this time with the algebra
$C_r$ but taking {\it one-half} the
highest root of $C_r$ to make the extension
(see \cite{Kac} formula in paragraph 6.4, bottom of
page 84). The formulas for
$\alpha_0$ and $\alpha_{-1}$ are $2 \alpha_0 = u - \theta$
and $2 \alpha_{-1} = -u -v $ (where $\theta$ is now the
highest root of $C_r$).  The Dynkin diagram of $A_{2 r}^{(2)'\wedge}$
is (Langlands) dual to that of $A_{2 r}^{(2)\wedge}$.
[Duality amounts to reversing the arrows in the Dynkin diagram, i.e., to
replacing the (generalized) Cartan matrix by its transpose.]

The algebras $A_{2 r}^{(2)\wedge}$ and $A_{2 r}^{(2)'\wedge}$ have rank
$r+2$ and are hyperbolic for $r \leq 4$.
The intermediate affine algebras are in all cases the twisted affine
algebras $A_{2 r}^{(2)}$.  By coupling to 3-dimensional gravity a coset
model $G/H$ where the restricted root system of the (real) Lie
algebra of the Lie group $G$ is of $bc_r$-type, one can realize all
the $A_{2r}^{(2)\wedge}$ algebras.

\subsubsection*{Twisted overextensions associated with
the highest short root}
We denote by $\theta_s$ the unique short root of heighest
weight.  It exists only for non-simply laced algebras and
has length $1$ (or $2/3$ for $G_2$).
The twisted overextensions are defined as the standard overextensions
but one uses instead the highest short root $\theta_s$.  The formulas
for the affine and overextended roots are
$$  \alpha_0 = u -  \theta_s , \; \; \;  \alpha_{-1} = - u - \frac{1}{2} v,
\; \; \; ({\cal L} = B_r, C_r, F_4) $$
or
$$  \alpha_0 = u -  \theta_s , \; \; \; \alpha_{-1} = - u - \frac{1}{3}v,
\; \; \; ({\cal L} = G_2). $$
[We choose the overextended root to have the same length as the affine
root and to be attached to it with a single link. This choice is
motivated by considerations of simplicity and yields the fourth
rank ten hyperbolic algebra when ${\cal L} = C_8$.]

The affine extensions generated by $\alpha_0, \cdots, \alpha_r$ are
respectively the twisted affine algebras
$D_{r+1}^{(2)}$ (${\cal L} = B_r$), $A_{2r-1}^{(2)}$
(${\cal L} = C_r$), $E_6^{(2)}$ (${\cal L} = F_4$) and
$D_4^{(3)}$ (${\cal L} = G_2$).  The overextensions $D_{r+1}^{(2)\wedge}$
have rank $r + 2$ and are hyperbolic for $r \leq 4$.  The overextensions
$A_{2r-1}^{(2) \wedge}$ have rank $r + 2$ and are hyperbolic for $r \leq
8$.  The last hyperbolic case, $r=8$, yields the algebra $A_{15}^{(2)
\wedge}$ also denoted $CE_{10}$.  It is the fourth rank-10
hyperbolic algebra, besides $E_{10}$, $BE_{10}$ and $DE_{10}$.
[$CE_{10}$ is also considered in \cite{HJK}.]
The overextensions $E_6^{(2)\wedge}$ (rank 6) and $D_4^{(3)\wedge}$
(rank 4) are hyperbolic.

\break
\subsubsection*{Dynkin diagrams}
We list below the Dynkin diagrams of all twisted overextensions.

\begin{figure}[ht]
\centerline{\includegraphics[scale=0.6]{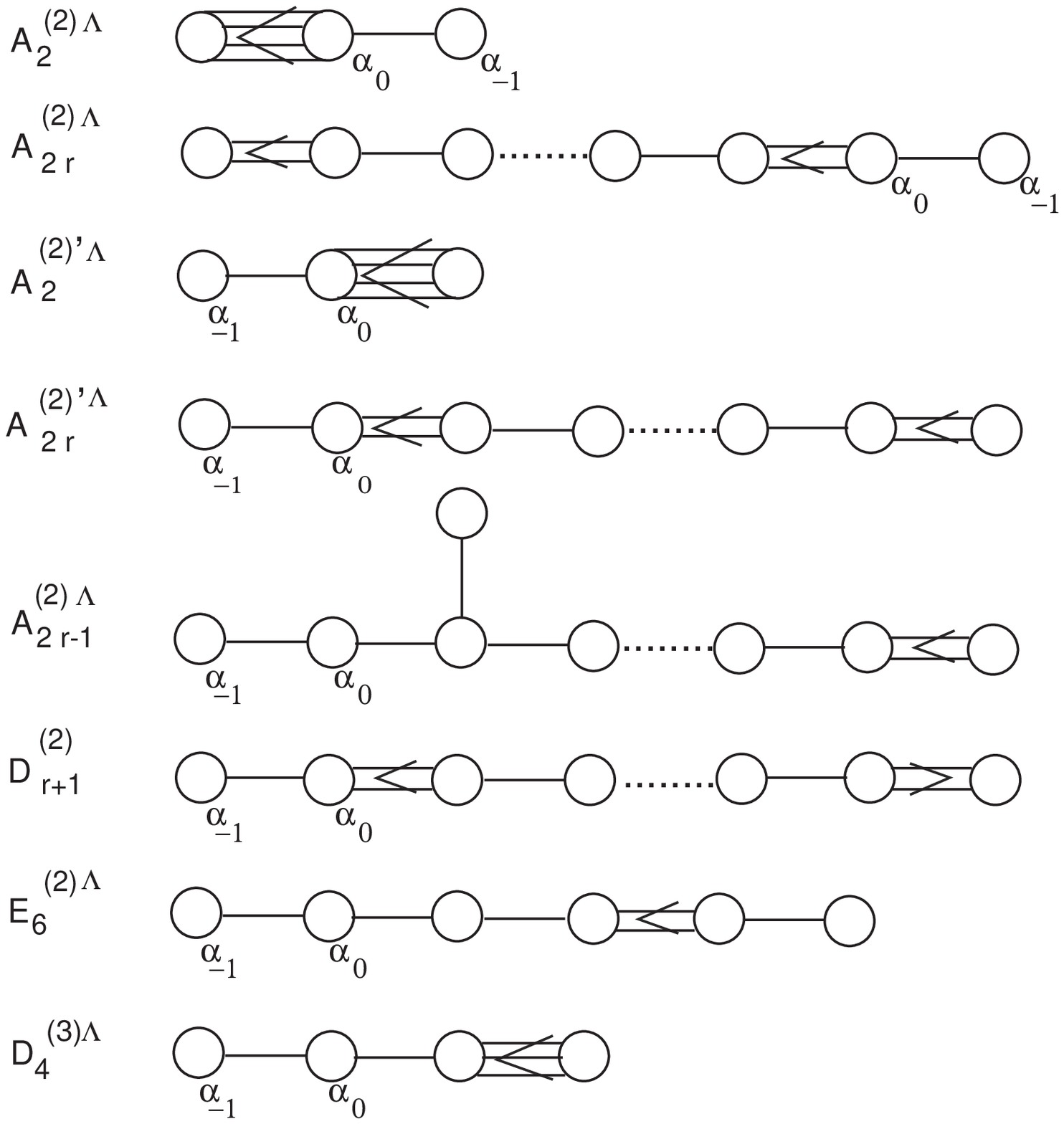}}
\end{figure}

\noindent
A satisfactory feature of the class of overextensions
(standard {\em and}
twisted) is that it is closed under duality.
For instance, $A_{2r-1}^{(2)\wedge}$ is dual to
$B_r^{\wedge \wedge}$.
In fact, one could get the twisted overextensions associated
with the highest short root from the standard overextensions
precisely by requiring closure
under duality.  A similar feature already holds
for the affine algebras.

\end{document}